\documentclass[aps,twocolumn,pra,superscriptaddress,showpacs]{revtex4}

\usepackage{mathrsfs}
\usepackage{amssymb}
\usepackage{amsmath}
\usepackage{graphicx}
\usepackage{epsfig}
\usepackage{txfonts}
\usepackage{subfigure}
\usepackage{amsfonts}
\usepackage{color}
\usepackage[colorlinks,citecolor=blue]{hyperref}
\usepackage{bm}

\begin{document}
	
	\title{Enhancing the sensitivity of quantum fiber-optical gyroscopes via a non-Gaussian-state probe}
	\author{Wen-Xun Zhang}
	\affiliation{Key Laboratory of Low-Dimensional Quantum Structures and Quantum Control of Ministry of Education, Synergetic Innovation Center for Quantum Effects and Applications,  XJ-Laboratory and Department of Physics, Hunan Normal University, Changsha 410081, China}
	
	\author{Rui Zhang}
	\affiliation{Key Laboratory of Low-Dimensional Quantum Structures and Quantum Control of Ministry of Education, Synergetic Innovation Center for Quantum Effects and Applications,  XJ-Laboratory and Department of Physics, Hunan Normal University, Changsha 410081, China}
	
	\author{Yunlan Zuo}
	\affiliation{Key Laboratory of Low-Dimensional Quantum Structures and Quantum Control of Ministry of Education, Synergetic Innovation Center for Quantum Effects and Applications,  XJ-Laboratory and Department of Physics, Hunan Normal University, Changsha 410081, China}
	
	\author{Le-Man Kuang}\email{lmkuang@hunnu.edu.cn}
	\affiliation{Key Laboratory of Low-Dimensional Quantum Structures and Quantum Control of Ministry of Education, Synergetic Innovation Center for Quantum Effects and Applications,  XJ-Laboratory and Department of Physics, Hunan Normal University, Changsha 410081, China}
	\affiliation{Academy for Quantum Science and Technology, Zhengzhou University of Light Industry, Zhengzhou 450002, China}
	
	
	\begin{abstract}
		
		We propose  a theoretical scheme to enhance the sensitivity of a quantum fiber-optical gyroscope (QFOG) via a non-Gaussian-state probe based on quadrature measurements of the optical field. The  non-Gaussian-state probe utilizes the product state comprising a photon-added coherent state (PACS) with photon excitations and a coherent state CS. We study the sensitivity of the QFOG,  and find that it can be significantly enhanced through increasing the photon excitations in the PACS probe. We investigate the influence of photon loss on the performance of QFOG and demonstrate that the PACS probe exhibits robust resistance to photon loss. Furthermore, we compare the performance of the QFOG using the PACS probe against two Gaussian-state probes: the CS probe and the squeezed state (SS) probe and indicate that the PACS probe offers a significant advantage in terms of sensitivity, regardless of photon loss, under the constraint condition of the same total number of input photons. Particularly, it is found that the sensitivity of the PACS probe can be three orders of magnitude higher than that of two Gaussian-state probes for certain values of the measured parameter. The capabilities of the non-Gaussian state probe on enhancing the sensitivity and resisting photon loss  could have a wide-ranging impact on future high-performance QFOGs.
		
	\end{abstract}
	
	\maketitle
	
	\section{\label{level}Introduction}
	
	Enhancing the precision of measured parameters remains a fundamental topic in quantum metrology and quantum sensing \cite{1Giovannetti,2Giovannetti,3Giovannetti,4Apellaniz,5Barbieri,6Degen,7Pirandola}. Nonclassical effects of quantum systems \cite{8Kwon} can be used to improve the performance of various sensing systems beyond the shot noise limit that applies to classical sources. Quantum-enhanced metrology and sensing study how to exploit quantum resources to enhance the sensitivity of parameter estimations. It has already been demonstrated that squeezed states \cite{9Caves,10LIGO,11Aasi,12LIGO,13Goodwin,14Ganapathy,15Anisimov,16Schnabel,17Zuo,18Zhao,19Pezz,20Nielsen}, entangled states
	\cite{21Zhang,22Wasilewski,23Zhao,24Belliardo,25Hao,26Boto,27Kok,28Joo,29Chen,30Maccone,31Li,32Zhuang,33Lloyd,34tTan,35Gallego,36Giovannetti,37Lee,38Joo,39Dowling}, and quantum phase transitions \cite{40Chen,41Chu,42Zhang,43Chu,44Lu,45Xu,46Jing,47Peng} are important quantum resources that offer quantum advantages in many emerging sensing applications.
	
	Recently, much attention has been paid to non-Gaussian states with Wigner negativity \cite{48Walschaers,49Ra,50Melalkia} due to theoretical and experimental progress in continuous-variable quantum information theory. Experimentally realizable non-Gaussian states include photon-added coherent (PAC) states  \cite{51Agarwal,52Zavatta,52Viciani,52Barbieri},  Schr\"{o}dinger cat states \cite{53Ourjoumtsev,54Bao,55Wang,56Hacker,57Li,58Liu,59Huang,60Zou,61Liao,62Han,63Qin},  and so on.
	Non-Gaussian states  are widely applied in quantum information processing, such as quantum error correction \cite{64Li,65Ofek,66Ma,67Schlegel}, quantum metrology  and quantum sensing \cite{28Joo,29Chen,68Zhang,69Lee,70Zurek,71Gessner,72Tatsuta}.

	As is well known, the fiber-optic gyroscope (FOG)  offers a high precision, compact solution for precision navigation in GPS denied environments, ultraprecise platform stabilization,
	and other inertial sensing applications \cite{75Lefevre}.
	The Quantum fiber optic gyroscope (QFOG) is a kind of quantum gyroscope that utilizes quantum resources such as quantum squeezing, quantum entanglement, and the non-Gaussianity of quantum states. \cite{76Lenef,77Gustavson,78Wu2007,79Canue,80Tackmann,81Stockton,82Savoie,83Dutta,84Avinadav,85Zhao,86Krzyzanowska,87Li2021,88Li2018,89Jiao,90Likai,91Li2023,92LiG2022,93Davuluri}. The QFOG aims to realize rotation sensing with high precision by quantum effects. To date, however, studies  on QFOGs have mainly focused on utilizing Gaussian states as quantum probes to improve the performance of the QFOG.
	
	M. Mehmet and coworkers \cite{94Mehmet} demonstrated that the nonclassical sensitivity of the QFOG was improved by 4.5 dB beyond the shot-noise limit by the injection of squeezed light, and showed that the quantum enhancement of local area (kilometer-size) optical fiber networks and fiber-based measurement devices is feasible with squeezed light.
	Inspired by distributed quantum sensing schemes \cite{95Zhuang,96Jacobs,97Proctor}, M. R. Grace et al. \cite{98Grace}  proposed  an entanglement-enhanced QFOG that employs a stacked
	array of multiple identical Mach-Zehnder interferometers, with one part of a multimode-entangled squeezed vacuum state distributed to each.

	In this paper, We propose  a theoretical scheme to enhance the sensitivity of a quantum fiber-optical gyroscope (QFOG) via a non-Gaussian-state probe based on quadrature measurements of the optical field. The  non-Gaussian-state probe utilizes the product state comprising a photon-added coherent state (PACS) with photon excitations and a coherent state CS. We study the sensitivity of the QFOG,  and find that it can be significantly enhanced through increasing the photon excitations in the PACS probe. We investigate the influence of photon loss on the performance of QFOG and demonstrate that the PACS probe exhibits robust resistance to photon loss. Furthermore, we compare the performance of the QFOG using the PACS probe against two Gaussian-state probes: the CS probe and the squeezed state (SS) probe and indicate that the PACS probe offers a significant advantage in terms of sensitivity, regardless of photon loss, under the constraint condition of the same total number of input photons. Particularly, it is found that the sensitivity of the PACS probe can be three orders of magnitude higher than that of two Gaussian-state probes for certain values of the measured parameter. The capabilities of the non-Gaussian state probe on enhancing the sensitivity and resisting photon loss  could have a wide-ranging impact on future high-performance QFOGs.

	This paper is structured as follows. In Sec. II, we introduce the quadrature-measurement-based QFOG scheme and present a general expression to calculate its sensitivity. In Sec. III, we analyze the sensitivity of the QFOG when the two-mode probe  is in a non-Gaussian state, which is a product state of  a  photon-added coherent state and a coherent state. In Sec. IV, we compare the sensitivity of the non-Gaussian-state probe with that of Gaussian-state probes and demonstrate the advantage of the non-Gaussian-state probe over the Gaussian-state probe under the same quantum resource conditions. Finally, our conclusions are summarized in Sec. V.

	\section{\label{level2}  Quadrature-measurement-based QFOG}

	In this section, we describe the quadrature-measurement-based QFOG scheme and present a general expression to calculate the QFOG sensitivity.
	The QFOG under our consideration uses a Sagnac interferometer to measure angular velocity via the optical path delay induced in counterpropagating paths
	around a rotating loop as shown in Fig. 1.  It consists of a  lossless 50:50 beam splitter (BS) and a rotating optic-fiber loop.  The quantum probe is the light state illuminating the gyroscope.
	Within the interferometer, the system is made of two counterpropagating modes $c$ and $d$, that is the output modes of  the BS.  The output modes of the BS  couple with the input modes through the following BS transformation
	\begin{equation}
		\begin{split}
			\left[ {\begin{array}{*{20}{c}}
					{{d}}\\
					{{c}}
			\end{array}} \right]& = \frac{1}{{\sqrt 2 }}\left[ {\begin{array}{*{20}{c}}
					1&{ - 1}\\
					1&1
			\end{array}} \right]\left[ {\begin{array}{*{20}{c}}
					{{a}}\\
					{{b}}
			\end{array}} \right] \\
		\end{split}
	\end{equation}
	
	The propagation of the output modes  of the BS   within the fiber can be described by the unitary operator
	\begin{equation}
		\begin{split}
			\left[ {\begin{array}{*{20}{c}}
					{{c'}}\\
					{{d'}}
			\end{array}} \right]& = \left[ {\begin{array}{*{20}{c}}
					0&{e^{-i\phi}}\\
					e^{i\phi}&0
			\end{array}} \right]\left[ {\begin{array}{*{20}{c}}
					{{d}}\\
					{{c}}
			\end{array}} \right] \\
		\end{split}
	\end{equation}
	where the phase $\phi$ is the Sagnac phase. It is proportional to the measured angular velocity  with relationship $\phi=T\Omega/2$ where the  scale factor $T$ is dependent on the optical center frequency of the laser,  the number of fiber loops in the coil, the directed area of a single fiber loop.

	The quadrature measurement will be made at the output modes $a'$ and $b'$ of the QFOG. These are related to the modes within the interferometer
	through the same  BS  performing the inverse transformation  in Eq. (1)
	
	\begin{equation}
		\begin{split}
			\left[ {\begin{array}{*{20}{c}}
					{{a'}}\\
					{{b'}}
			\end{array}} \right]& = \frac{1}{{\sqrt 2 }}\left[ {\begin{array}{*{20}{c}}
					1&{ 1}\\
					-1&1
			\end{array}} \right]\left[ {\begin{array}{*{20}{c}}
					{{c'}}\\
					{{d'}}
			\end{array}} \right] \\
		\end{split}
	\end{equation}
	
	Making use of Eqs. (1), (2), and (3), we can obtain the  transformation relation between the input-output modes of the QFOG
	\begin{equation}
		\begin{split}
			\left[ {\begin{array}{*{20}{c}}
					{{a'}}\\
					{{b'}}
			\end{array}} \right]& = \left[ {\begin{array}{*{20}{c}}
					{\cos \phi }&{ - i\sin \phi }\\
					{i\sin \phi }&{ - \cos \phi }
			\end{array}} \right]\left[ {\begin{array}{*{20}{c}}
					{{a}}\\
					{{b}}
			\end{array}} \right]
		\end{split}
	\end{equation}

	\begin{figure}[t]
		\centerline{
			\includegraphics[width=0.48\textwidth]{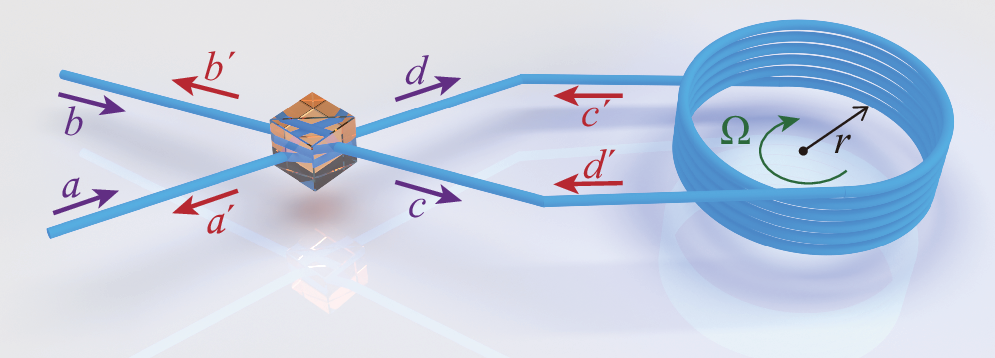}}
		\caption{Schematic of the quantum fiber-optic gyroscope.} \label{fig1}
	\end{figure}

	Assuming symmetric loss processes that impose equal transmissivities $\gamma$ on the two opposing paths around
	the fiber coil, we model fiber loss with the pure-loss channel
	\begin{equation}
		{\Phi _\gamma }\left(\left[ {\begin{array}{*{20}{c}}
				{{a}}\\
				{{b}}
		\end{array}} \right]\right) = \left[ {\begin{array}{*{20}{c}}
				{\sqrt \gamma  {a} + \sqrt {1 - \gamma } {e_a}}\\
				{\sqrt \gamma  {b} + \sqrt {1 - \gamma } {e_b}}
		\end{array}} \right]
	\end{equation}
	where $e_a$  and $e_b$ are  two ancillary vacuum modes to describe photon loss in the two probe modes $a$  and $b$ of the QFOG, respectively. The parameter of modelling photon loss takes $0\leq\gamma\leq 1$. $\gamma=1$ implies no photon loss while $\gamma=0$ denotes complete photon loss.
	
	From Eqs. (4) and (5), we can obtain the transformation relation between the input and output modes of the QFOG in the presence of  photon loss
	\begin{eqnarray}
		a'&=&( {\sqrt \gamma a + \sqrt {1 - \gamma } {e_a}})\cos \phi \nonumber \\
		&& - i({\sqrt \gamma b + \sqrt {1 - \gamma } {e_b}})\sin \phi \\
		b'&=&i({\sqrt \gamma a + \sqrt {1 - \gamma } {e_a}})\sin \phi  \nonumber \\
		&& - ({\sqrt \gamma  b + \sqrt {1 - \gamma } {e_b}})\cos \phi
	\end{eqnarray}

	In what follows, we investigate the sensitivity of the QFOG by using the quadrature measurement. For a boson field with the annihilation and creation operators $f$ and $f^{\dagger }$ with the commutation relation $[f, f^{\dagger}]=1$, the quadrature operators are defined as
	\begin{equation}
		X_{1f}=\frac{1}{2}(f+f^{\dagger }), \hspace{0.5cm}
		X_{2f}=\frac{1}{2i}(f-f^{\dagger })
	\end{equation}
	which satisfy the commutation relation $[X_{1f}, X_{2f}]=i/2$.
	
	For the output mode $b'$, making use of Eq. (7), we can obtain its  quadrature operators
	\begin{align}
		X_{1b'}=&-(\sqrt{\gamma }X_{2a}+\sqrt{1-\gamma }X_{2ea})\sin \phi \nonumber \\
		& -(\sqrt{\gamma }X_{1b}+\sqrt{1-\gamma }X_{1eb})\cos \phi \\
		X_{2b'}=&(\sqrt{\gamma }X_{1a}+\sqrt{1-\gamma }X_{1ea})\sin \phi \nonumber \\
		&-(\sqrt{\gamma }X_{2b}+\sqrt{1-\gamma }X_{2eb})\cos \phi
	\end{align}
	
	Assuming that the two ancillary modes describing photon loss are in a vacuum state, we can obtain the variances of the quadrature operators of the $b'$ mode from Eqs. (9) and (10)
	\begin{eqnarray}
		\Delta X_{1b'}^{2}  &=&\gamma [ \Delta X_{2a}^{2}\sin
		^{2}\phi + \Delta X_{1b}^{2}  \cos ^{2}\phi ]+ \frac{1-\gamma }{4} \\
		\Delta X_{2b'}^{2}  &=&\gamma [ \Delta X_{1a}^{2}\sin
		^{2}\phi + \Delta X_{2b}^{2}  \cos ^{2}\phi ]+ \frac{1-\gamma }{4}
	\end{eqnarray}
	where the quadrature variances appeared in the right-hand side of above equations are calculated in the input states of the input modes $a$ and $b$. In the derivation process of the variances of the quadrature operators of the $b'$ mode, we have used the fact that the mean values of the quadrature operators of the two ancillary modes in  a vacuum state vanish, and their variances are equal to $1/4$.
	
	The sensitivity of the QFOG  based on the quadrature $X_{ib'}$ measurement  can be expressed from  the error transfer equation
	\begin{equation}
		\Delta \Omega ^{2}_{X_{ib'}}=\frac{\Delta X_{ib'}^{2}}{|\partial \langle X_{ib'}\rangle/\partial\Omega|^2},  \hspace{0.5cm }  (i=1, 2)
	\end{equation}
	
	Making use of the relation  $\phi=T\Omega/2$ and Eqs. (9)-(12) we can find that the sensitivity of the QFOG is given by
	\begin{eqnarray}
		\Delta \Omega ^{2}_{X_{1b'}}  &=&\frac{1}{T^{2}}\frac{4\left(\Delta X_{2a}^{2}
			\sin ^{2}\phi + \Delta X_{1b}^{2}\cos^{2}\phi \right)+\frac{1-\gamma }{\gamma }}
		{\left[\left\langle X_{2a}\right\rangle \cos\phi -\left\langle X_{1b}\right\rangle\sin\phi\right]^2} \\
		\Delta \Omega ^{2}_{X_{2b'}}  &=&\frac{1}{T^{2}}\frac{4\left(\Delta X_{1a}^{2}
			\sin ^{2}\phi + \Delta X_{2b}^{2}\cos^{2}\phi \right)+\frac{1-\gamma }{\gamma }}
		{\left[\left\langle X_{1a}\right\rangle\cos\phi +\left\langle X_{2b}\right\rangle\sin\phi\right]^2}
	\end{eqnarray}

	\section{\label{level3} The sensitivity of the QFOG with non-Gaussian-state probe}
	
	In this section, we study the sensitivity  of the QFOG when the probe mode $a$ is in a  photon-added coherent state $\left\vert \psi\right\rangle_a $ while the probe mode $b$ is in a coherent state $\left\vert \psi\right\rangle_b=|\beta\rangle$. We show that the sensitivity  of the QFOG can be enhanced through increasing the number of photon excitations in the photon-added coherent state.
	
	We consider a photon-added coherent state with $m$ excitations defined by
	\begin{equation}
		\left\vert \psi\right\rangle_a =N_{m}a^{\dagger m}\left\vert \alpha \right\rangle
	\end{equation}
	where  $N_{m}$ is the normalization constant given by
	\begin{equation}
		N_{m} =\left[m!L_{m}(-\left\vert \alpha \right\vert ^{2})\right]^{-1/2}
	\end{equation}
	with $L_{m}(x)$ being the Laguerre polynomial defined by
	\begin{equation}
		L_{m}(x)=\sum_{k=0}^{m}(-1)^{k}\frac{m!}{(k!)^{2}(m-k)!}x^{k}
	\end{equation}
	
	We take the quadrature measurement of $X_{2b'}$ for the output mode of the QFOG. Assuming that both $\alpha$ and $\beta$ are real numbers, from Eq. (15)  we can find that the sensitivity of the QFOG is given by
	\begin{equation}
		\Delta \Omega ^{2}_{NG}=\frac{1}{T^{2}}\frac{\left(4\Delta X_{1a}^{2}-1\right)\sin ^{2}\phi +\frac{1}{\gamma}}{\left(\left\langle X_{1a}\right\rangle\cos \phi\right)^{2}}
	\end{equation}
	where the mean value and variance of the quadrature operator $X_{1a}$ in the PACS given by Eq. (16) are given by
	\begin{align}
		\left\langle X_{1a}\right\rangle=&\frac{L_{m}^{1}(-\alpha ^{2})}{L_{m}(-\alpha ^{2})}\alpha\\
		\Delta X_{1a}^{2}=&\frac{(m+1)L_{m+1}(-\alpha ^{2})+\alpha ^{2}L_{m}^{2}(-\alpha ^{2})}{2L_{m}(-\alpha ^{2})}\nonumber\\
		&-\alpha ^{2}\frac{[L_{m}^{1}(-\alpha ^{2})]^{2}}{[L_{m}(-\alpha ^{2})]^{2}}-\frac{1}{4}
	\end{align}
	where $\mu$-th order associated Laguerre polynomial defined by
	\begin{equation}
		L_{m}^{\mu }(x)=\sum_{k=0}^{m}(-1)^{k}\frac{(\mu +m)!}{k!(m-k)!(\mu +k)!}x^{k}
	\end{equation}
	
	From Eq. (19) we can observe that the sensitivity of the QFOG  is independent of the state of the input mode $b$. Then, we can take  vacuum state as the state of the input mode $b$  to save quantum resource. Hence, the total number of photons in the input state of the QFOG is equal to that in the input mode $a$ of the PACS given by
	\begin{align}
		\overline{n}_a=& _a\left\langle \psi|a^{\dagger }a|\psi\right\rangle_a \nonumber\\
		=&\frac{(m+1)L_{m+1}(-\left\vert \alpha \right\vert ^{2})}{L_{m}(-\left\vert \alpha \right\vert ^{2})}-1
	\end{align}
	
	However, the situation will change when we take $\alpha$ as  a real number, and $\beta=iy$   a pure imaginary number ($y$ is real). When the input state of the PACS probe is $| \psi\rangle_a\otimes|\psi\rangle_b=|PACS\rangle\otimes|iy\rangle$,  the sensitivity of the QFOG (15)  is reduced to
	\begin{equation}
		\Delta \Omega ^{2}_{NG}=\frac{1}{T^{2}}\frac{\left(4\Delta X_{1a}^{2}-1\right)\sin ^{2}\phi +\frac{1}{\gamma}}
		{\left(\left\langle X_{1a}\right\rangle\cos \phi +y\sin \phi\right)^{2}}
	\end{equation}
	which indicates that the sensitivity of the QFOG depends on state parameters of the two input modes, $a$ and $b$.

	From Eqs.  (19) and (24)  we can also see that the smaller the photon loss rate $\gamma$,  the larger the value of the sensitivity  $\Delta \Omega ^{2}_{NG}$. Hence, the sensitivity of the QFOG deteriorates with the increase of photon loss. In the absence of photon loss ($\gamma=1$), the  QFOG reaches the best sensitivity.
	
	It is interesting to note that  in the small rotation regime of $\phi\ll 1$, the QFOG has the same  sensitivity for two different probe states $|PACS\rangle|0\rangle$ and $|PACS\rangle|iy\rangle$.
	In fact, we have $\sin \phi\approx 0$ and $\cos \phi\approx 1$ in the small rotation regime of $\phi\ll 1$,  hence  the sensitivity of the QFOG given by Eqs. (19) and (24) becomes the same expression
	\begin{equation}
		\Delta \Omega ^{2}_{NG}\approx   \frac{1}{T^{2}\gamma\left\langle X_{1a}\right\rangle^{2}}
	\end{equation}
	which implies that the sensitivity of the QFOG is independent of the measured parameter  in the  small rotation regime.

	\begin{figure}[t]
		\centerline{
			\includegraphics[width=0.48\textwidth]{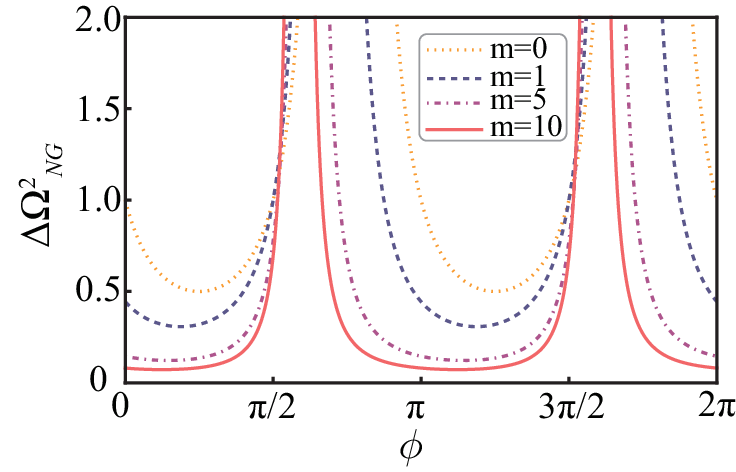}}
		\caption{The sensitivity of the QFOG with respect to the Sagnac phase  in the absence of  photon loss  ($\gamma=1$). The sensitivity is scaled by $1/T^2$. The dot, dash, dash-dot, and  solid lines correspond to $m=0, 1, 5,$ and $10$, respectively.}
	\end{figure}
	
	We now turn to the numerical investigation of the sensitivity of the QFOG. We only study the sensitivity of the QFOG  based on the quadrature $X_{2b'}$ measurement  given by Eq. (15) since the numerical results in next section indicate that the sensitivity of this case can surpass that of Gaussian-state probe. In Fig. 2 we plot the sensitivity of the QFOG with respect to the Sagnac phase  in the absence of  photon loss  ($\gamma=1$) when we take $\alpha=1$ and $y=1$ for different  excitation number $m=0, 1, 5,$ and $10$, respectively. Here the sensitivity is scaled by $1/T^2$. From Fig. 2 we can see that (i) the sensitivity of the QFOG periodically changes with respect to the Sagnac phase. (ii) The QFOG can show high performance in most regime of one phase period. (iii) The values of the sensitivity $\Delta \Omega ^{2}_{NG}$ decrease with the increase of the excitation number $m$. Hence the performance of the QFOG is improved with  the increase of the excitation number.
	
	In Fig. 3 we plot the sensitivity of the QFOG with respect to the Sagnac phase  in the presence of  photon loss  ($\gamma=0.5$) when we take $\alpha=1$ and $y=1$ for different  excitation number $m=0, 1, 5,$ and $10$, respectively. From Fig. 3 we can find  that quantum excitations of a coherent state can enhance the sensitivity of the QFOG even though there exists photon loss.
	
	In order to further observe the influence of photon loss on the sensitivity, in Fig. 4 we show the sensitivity of the QFOG with respect to the  photon loss when we  take   $\phi=\pi/4$, $\alpha=1$,  and $y=1$ for different  excitation number $m=0, 1, 5,$ and $10$, respectively. From  Fig. 4, we can see that the probe state with the larger excitation number exhibits the stronger ability against photon loss.

	\begin{figure}[t]
		\centerline{
			\includegraphics[width=0.48\textwidth]{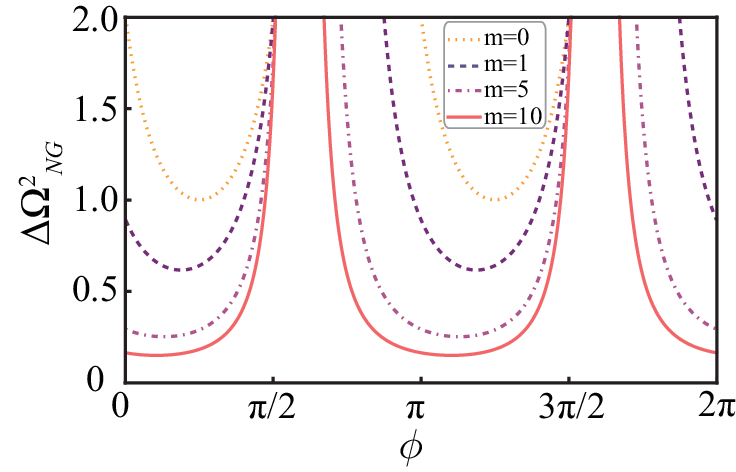}}
		\caption{The sensitivity of the QFOG with respect to the Sagnac phase  in the presence of  photon loss  ($\gamma=0.5$). The sensitivity is scaled by $1/T^2$. The dot, dash, dash-dot, and  solid lines correspond to $m=0, 1, 5,$ and $10$, respectively.}
	\end{figure}

	\begin{figure}[t]
		\centerline{
			\includegraphics[width=0.48\textwidth]{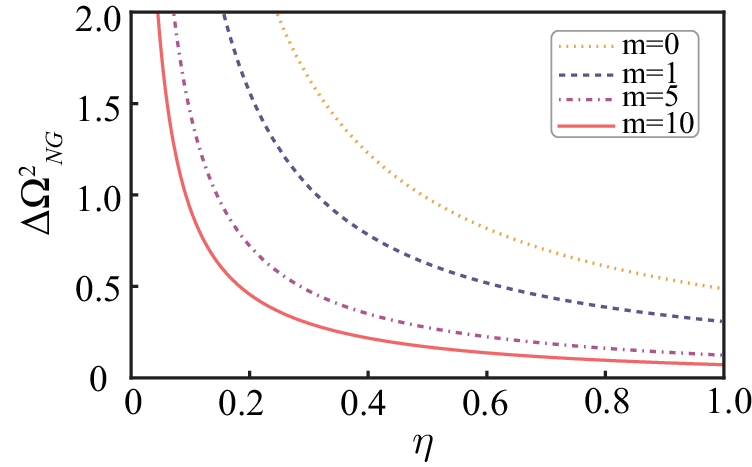}}
		\caption{The sensitivity of the QFOG   with respect to the  photon loss when the  Sagnac phase takes  $\phi=\pi/4$. The sensitivity is scaled by $1/T^2$.  The dot, dash, dash-dot, and  solid lines correspond to $m=0, 1, 5,$ and $10$, respectively.}
	\end{figure}

	\section{\label{level4} Comparison with  the sensitivity  of the QFOG with Gaussian-state probe}
	
	In this section, we compare the sensitivity  of  the non-Gaussian-state probe with that of  Gaussian-state probes to show the advantage of the non-Gaussian-state probe over the Gaussian-state probe under the same condition of  the quantum resource. We will consider two types of Gaussian-state probes. We regard  the first Gaussian-state probe as the CS probe in which the input state of the probe is product of two coherent state $|\psi _{a}\rangle\otimes |\psi _{b}\rangle=|\alpha_c\rangle\otimes|iy\rangle$ with $\alpha_c$ and $y$ being real. The second Gaussian-state probe is the squeezed-state probe in which the input state of the probe is product of a coherent state and a squeezed state $|\psi _{a}\rangle\otimes |\psi _{a}\rangle_b=|\alpha_c\rangle\otimes| \xi\rangle$ with $\xi=-r$.

	In order to study the advantage of the non-Gaussian  probe over the Gaussian probe in the QFOG under the condition of  the same input photon number, we introduce the ratio
	\begin{equation}
		R =\frac{\Delta \Omega ^{2}_{NG}}{\Delta \Omega ^{2}_{G}}
	\end{equation}
	where $\Delta \Omega ^{2}_{NG}$ and $\Delta \Omega ^{2}_{G}$ are the sensitivity of the QFOG when the input state are non-Gaussian  and Gaussian states, respectively.
	For the non-Gaussian  probe with the input state $|\psi _{a}\rangle\otimes |\psi _{b}\rangle=|PACS\rangle\otimes|iy\rangle$, the sensitivity of the QFOG  based on the quadrature $X_{2b'}$ measurement is given by Eq. (24). Therefore, $R<1$ means that the non-Gaussian  QFOG has better sensitivity than the Gaussian  QFOG.
	
	\subsection{\label{level4} Comparison with   the QFOG with coherent-state probe}
	
	For the  coherent-state probe in which  the input state of the QFOG is the product of two coherent states $|\psi _{a}\rangle\otimes |\psi _{b}\rangle=|\alpha_c\rangle\otimes|iy\rangle$, the sensitivity of the QFOG  based on the quadrature $X_{2b'}$ measurement can be obtained from  Eq. (24) by setting $m=0$ and replacing $\alpha$ with $\alpha_C$. We can find that the sensitivity of the QFOG with the CS probe is given by
	\begin{equation}
		\Delta \Omega ^{2}_{CS}=\frac{1}{ T^{2}}\frac{1}{\gamma (\alpha _{c}\cos \phi +y\sin \phi )^{2}}
	\end{equation}
	
	From the constraint condition that the PACS probe and the CS probe have the same input photon number, we can obtain the following relation between state parameters of the PACS and CS probes
	\begin{equation}
		\alpha _{c}^{2}=\frac{(m+1)L_{m+1}(-\alpha ^{2})}{L_{m}(-\alpha ^{2})}-1
	\end{equation}
	
	Making use of the sensitivity of the QFOG given by Eqs. (24) and (27), from Eq. (26) we can get the sensitivity ratio between the PACS and CS QFOG
	\begin{equation}
		R_{NG-CS}=\frac{\left[\gamma\left(4\Delta X_{1a}^{2}-1\right)\sin ^{2}\phi +1\right](\alpha _{c}\cos \phi +y\sin \phi )^{2}}
		{\left(\left\langle X_{1a}\right\rangle\cos \phi +y\sin \phi\right)^{2}}
	\end{equation}
	where $\left\langle X_{1a}\right\rangle$ and $\Delta X_{1a}^{2}$ are  the expectation value and variance of the first quadrature operator of the probe mode $a$ in the PACS with $m$ photon excitations, they are have calculated in Eqs. (20) and (21).

	\begin{figure}[t]
		\centerline{
			\includegraphics[width=0.48\textwidth]{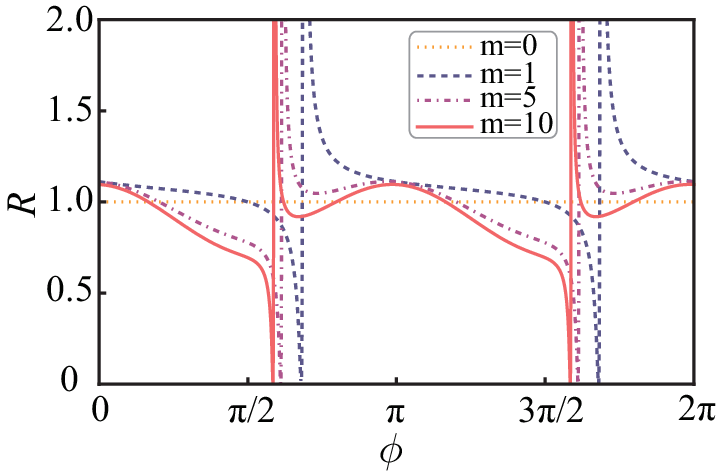}}
		\caption{The sensitivity ratio of the PACS and CS QFOG   with respect to the Sagnac phase under the constraint of the same total number of input photons  in the absence of  photon loss  ($\gamma=1$).  The dot, dash, dash-dot, and  solid lines correspond to $m=0, 1, 5,$ and $10$, respectively.}
	\end{figure}

	In Fig. 5, we show the sensitivity ratio of the PACS and CS QFOG  with respect to the Sagnac phase  under the constraint of the same total number of input photons in the absence of  photon loss  ($\gamma=1$)  for different  excitation numbers $m=0, 1, 5,$ and $10$, respectively.  Here we take the probe parameters $\alpha=1$ and $y=1$.
	From Fig. 5 we can see that (i) the sensitivity of the QFOG periodically varies with respect to the Sagnac phase. (ii) The value of $R$ is less than unit in most regime of one phase period. This implies that the performance of the PACS QFOG is better than that of the CS QFOG in wide regime of one phase period. (iii) The values of the sensitivity ratio $R$ decrease with the increase of the excitation number $m$. Hence  quantum advantage of the PACS QFOG over the CS QFOG is more pronounced  with  the increase of the excitation number. (iv) There exists a best phase point at which the sensitivity ratio can suddenly change, and the PACS probe is significantly better than  the CS probe. In fact, the further numerical analyse indicates that the sensitivity ratio of the PACS and CS QFOG can reach $R=0.0011$ at the phase $\phi=0.5844\pi$ for photon excitations $m=10$.
	
	In Fig. 6,  under the constraint of the same input mean photon number we plot the sensitivity ratio of the PACS and CS QFOG  with respect to the Sagnac phase in the presence of  photon loss  ($\gamma=0.5$)  for different  excitation numbers $m=0, 1, 5,$ and $10$, respectively.  Here we also take the probe parameters $\alpha=1$ and $y=1$.
	Comparing Fig. 6 with Fig. 5 we can find that photon loss generally leads to narrowing the advantage range of the PACS probe with respect to the CS probe.
	
	From the above numerical analyses we can conclude that  the PACS probe can exhibit stronger quantum advantage in the phase estimation over the CS probe whether in the absence or presence
	of photon loss under the constraint condition of the same input mean photon number.
	Firstly, the higher excitation PACS probe can create better phase sensitivity limit than the CS probe.
	Secondly, the PACS probe has stronger capability against the photon loss. In particular, the PACS probe is significantly better than  the CS probe at the best phase point.

	\begin{figure}[t]
		\centerline{
			\includegraphics[width=0.48\textwidth]{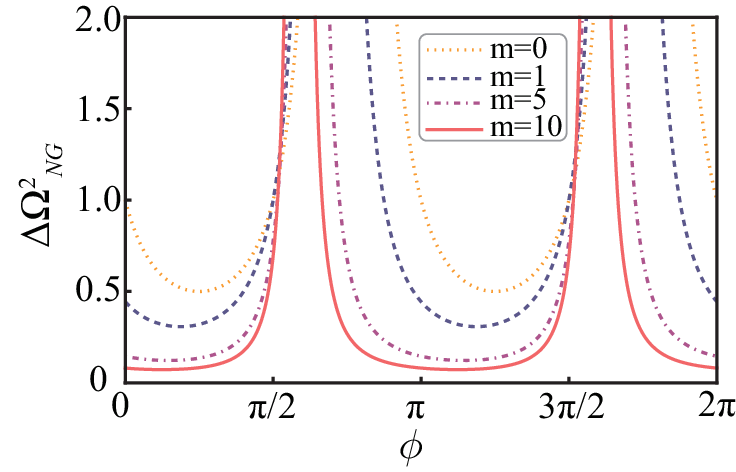}}
		\caption{The sensitivity ratio of the PACS and CS QFOG   with respect to the Sagnac phase under the constraint of the same total number of input photons  in the absence of  photon loss  ($\gamma=0.5$).  The dot, dash, dash-dot, and  solid lines correspond to $m=0, 1, 5,$ and $10$, respectively.}
	\end{figure}

	\subsection{\label{level4} Comparison with the QFOG with squeezed-state probe}
	
	In this subsection, we make a comparison on the sensitivity of the QFOP between the non-Gaussian probe, the PACS probe, and the second kind of Gaussian probe, the squeezed-state probe. Assuming that the input state of the  squeezed-state  probe is the product of a coherent state and a squeezed state $|\psi _{a}\rangle\otimes |\psi _{b}\rangle=|\alpha_c\rangle\otimes|\xi\rangle$ with the polarization decomposition of the squeezing parameter $\xi=re^{i\delta}$.
	
	Based on the quadrature $X_{2b'}$ measurement, the sensitivity of the QFOG with the SS probe   can be obtained from  Eq. (24) with the following expression
	\begin{equation}\label{yasuoh}
		\Delta \Omega ^{2}_{SS}=\frac{1}{T^{2}}\frac{\left(e^{-2r}-1\right)\cos ^{2}\phi +\frac{1}{\gamma }}{\alpha _{c}^{2}\cos^{2}\phi }
	\end{equation}
	where we have taken $\delta=\pi$ and used the following expect values and variances
	\begin{align}
		\left\langle X_{1a}\right\rangle =&\alpha_c, \hspace{0.5cm}  \Delta X_{1a}^{2} =\frac{1}{4} \\
		\left\langle X_{2,b}\right\rangle=&0, \hspace{0.5cm} \Delta X_{2,b}^{2} =\frac{1}{4}e^{-2r}
	\end{align}

	From the constraint condition that the PACS probe and the SS probe have the same input photon number, we can obtain the following relation between state parameters of the PACS and SS probes
	\begin{equation}
		\alpha _{c}^{2}=\frac{(m+1)L_{m+1}(-\alpha ^{2})}{L_{m}(-\alpha ^{2})}-1, \hspace{0.5cm}
		y^{2}=\sinh ^{2}r
	\end{equation}
	from which we can obtain $e^r=y + \sqrt{1+y^2}$. Hence the state parameters of the SS probe $(\alpha _{c}, r)$ can be expressed in terms of the state parameters of the PACS probe $(\alpha, m, y)$.

	Making use of the sensitivity of the QFOG given by Eqs. (24) and (30), from Eq. (26) we can get the sensitivity ratio between the PACS and SS QFOG
	\begin{equation}
		R_{NG-SS}=\frac{\left[\gamma\left(4\Delta X_{1a}^{2}-1\right)\sin ^{2}\phi +1\right]\alpha _{c}^{2}\cos^{2}\phi}
		{\left(\left\langle X_{1a}\right\rangle\cos \phi +y\sin \phi\right)^{2}\left[\gamma\left(e^{-2r}-1\right)\cos ^{2}\phi +1\right]}
	\end{equation}
	where $\left\langle X_{1a}\right\rangle$ and $\Delta X_{1a}^{2}$ are  the expectation value and variance of the first quadrature operator of the probe mode $a$ in the PACS with $m$ photon excitations, they have been calculated in Eqs. (20) and (21).

	\begin{figure}[t]
		\centerline{
			\includegraphics[width=0.48\textwidth]{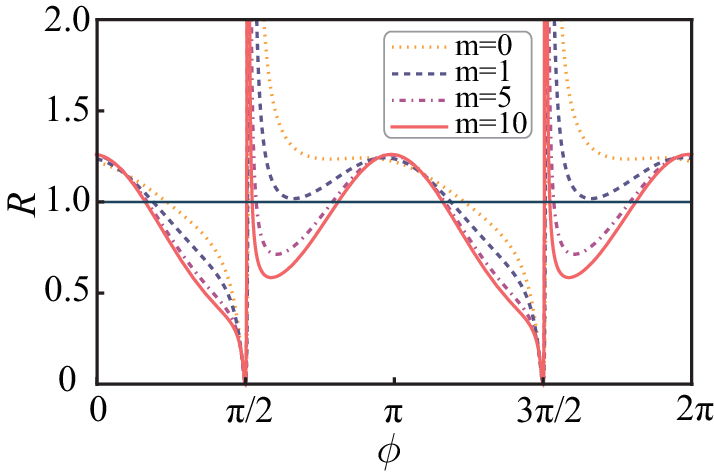}}
		\caption{The sensitivity ratio of the PACS and CS QFOG   with respect to the Sagnac phase under the constraint of the same total number of  input photons  in the absence of  photon loss  ($\gamma=1$).  The dot, dash, dash-dot, and  solid lines correspond to $m=0, 1, 5,$ and $10$, respectively.}
	\end{figure}

	\begin{figure}[t]
		\centerline{
			\includegraphics[width=0.48\textwidth]{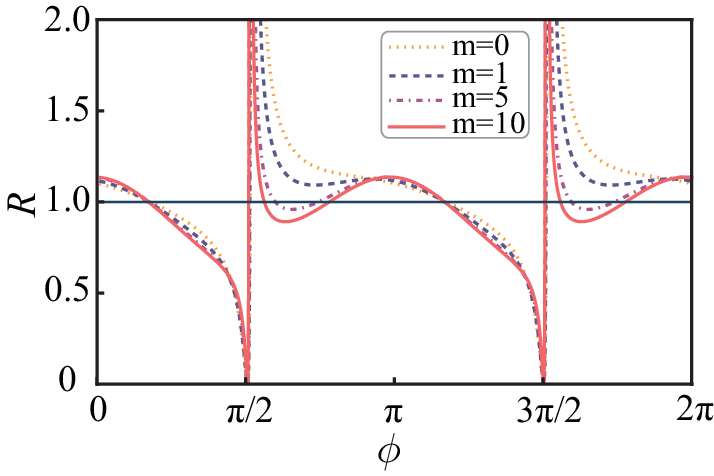}}
		\caption{The sensitivity ratio of the PACS and SS QFOG   with respect to the Sagnac phase under the constraint of the same total  number of  input photons  in the absence of  photon loss  ($\gamma=0.5$).  The dot, dash, dash-dot, and  solid lines correspond to $m=0, 1, 5,$ and $10$, respectively.}
	\end{figure}

	In Fig. 7, we show the sensitivity ratio of the PACS and SS QFOG  with respect to the Sagnac phase  under the constraint of the same total number of input photons in the absence of  photon loss  ($\gamma=1$)  for different  excitation numbers $m=0, 1, 5,$ and $10$, respectively.  Here we take the probe parameters $\alpha=1$ and $y=1$.
	Fig. 7 indicates that  when $m\geq1$, the sensitivity ratio $R$ is less than unit in the phase regime of $\pi/4 <\phi< \pi/2$, and the larger the photon-excitation number $m$, the smaller the sensitivity ratio $R$. This implies that the sensitivity of the PACS QFOG  is better than that of the SS QFOG, and the increase of the photon-excitation number can enhance the advantage of the  PACS QFOG over the  SS QFOG. When $m\geq 2$, we can find the second advantageous phase regime, $\pi/2 <\phi< 3\pi/4$, in which  the sensitivity of the PACS QFOG  is also better than that of the SS QFOG.
	
	From Fig. 7 we can see that $\phi=\pi/2$ is the transition point of the  sensitivity ratio at which the sensitivity ratio can suddenly change between $R<1$ and $R>1$. In the small neighborhood on the left-hand side of the transition point,  the PACS QFOG can exhibit the best advantage of  the sensitivity over  the SS QFOG.  For example, from Fig. 7 we can find the sensitivity ratio of the PACS and SS QFOG can reach $R=0.006$ at the phase point  $\phi=0.499\pi$ for photon excitations $m=10$.
	
	In Fig. 8,  we plot the sensitivity ratio of the PACS and CS QFOG  with respect to the Sagnac phase under the constraint of the same input mean photon number  in the presence of  photon loss  ($\gamma=0.5$)  for different  excitation numbers $m=0, 1, 5,$ and $10$, respectively.  Here we also take the probe parameters $\alpha=1$ and $y=1$.
	Comparing Fig. 8 with Fig. 7 we can find that photon loss suppress the advantage of the PACS-probe sensitivity with respect to the SS probe.
	
	\section{\label{level5}Conclusions}
	
	In this work, we have proposed a theoretical scheme based on the quadrature measurement of the optical field  to enhance the sensitivity of the QFOG via a non-Gaussian quantum probe. We constructed the QFOG quantum probe using the product state of the PACS with $m$-photon excitations and the CS. We have studied the sensitivity of the QFOG using the non-Gaussian quantum probe , which highlights some relevant features of the QFOG. We have found that  the sensitivity of the QFOG can be significantly enhanced by increasing the photon excitations in the PACS probe. We have also investigated the influence of photon loss on the performance of QFOG. It has been found that the probe state with higher excitation numbers exhibits stronger resistance to photon loss. The non-Gaussian-state offers two significant advantages over the Gaussian-state: enhanced sensitivity and robust resistance to photon loss. Furthermore, we have compared the performance of the QFOG using the PACS probe against the CS and SS probes by focusing on the sensitivity ratio. It has been demonstrated that the PACS probe can exhibit stronger quantum advantage in the sensitivity of  QFOG over the CS and SS probes regardless of photon loss under the constraint condition of the same  total number of input  photons. In particular, it has been indicated that the sensitivity of the PACS probe can be three orders of magnitude higher than that of two Gaussian-state probes for certain values of the measured parameter.
	The capabilities of the non-Gaussian state probe in enhancing the sensitivity and resisting photon loss compared to Gaussian state probes could have a wide-ranging impact on future high-performance QFOGs.

	\begin{acknowledgements}
		W. X. Zhang and R. Zhang are co-first authors with equal contribution to this work. L.-M. K. is supported by the Natural Science Foundation of China (NSFC) (Grant Nos. 12247105, 12175060, 11935006), Hunan provincial sci-tech program (Grant nos. 2020RC4047, 2023ZJ1010) and  XJ-Lab key project (Grant No. 23XJ02001). 
	\end{acknowledgements}
	
	%
	
	%
	
\end{document}